\documentclass[pra,twocolumn,superscriptaddress,amsmath,amssymb,aps,floatfix]{revtex4-2}
\usepackage{graphicx}
\usepackage{dcolumn}
\usepackage{bm}
\usepackage{amsmath,amssymb}
\usepackage{graphics}
\usepackage{amsfonts}
\usepackage{epsfig}
\usepackage{float}
\usepackage{hyperref}
\usepackage{adjustbox}
\usepackage{rotating}
\usepackage{float}
\usepackage{dblfloatfix}
\usepackage{blindtext}
\usepackage{multirow}
\usepackage{graphicx}
\usepackage{dcolumn}
\usepackage{bbold}
\usepackage{bm}
\usepackage{amsmath}
\usepackage{import}
\usepackage{physics}
\usepackage{verbatim}
\usepackage{listings}
\usepackage{xcolor}
\usepackage{xr-hyper}
\usepackage{hyperref}
\usepackage{cleveref}
\usepackage{newfloat}
\usepackage{caption}
\usepackage{subcaption}
\DeclareFloatingEnvironment[name={Fig S},fileext=lof]{suppfigure}
\begin{document}
\title{Designing three-way entangled and nonlocal two-way entangled single particle states via alternate quantum walks}
\author{Dinesh Kumar Panda}
\email{dineshkumar.quantum@gmail.com}
\author{Colin Benjamin}
\email{colin.nano@gmail.com}
\affiliation{School of Physical Sciences, National Institute of Science Education and Research Bhubaneswar, Jatni 752050, India}
\affiliation{Homi Bhabha National Institute, Training School Complex, Anushaktinagar, Mumbai
400094, India}
\begin{abstract}
Entanglement with single-particle states is advantageous in quantum technology because of their ability to encode and process information more securely than their multi-particle analogs. Three-way and nonlocal two-way entangled single-particle states are desirable in this context. Herein, we generate genuine three-way entanglement from an initially separable state involving three degrees of freedom of a quantum particle, which evolves via a 2D alternate quantum walk employing a resource-saving single-qubit coin. We achieve maximum possible values for the three-way entanglement quantified by the $\pi$-tangle between the three degrees of freedom. We also generate optimal nonlocal two-way entanglement, quantified by the negativity between the nonlocal position degrees of freedom of the particle. This prepared architecture using quantum walks can be experimentally realized with a photon.
\end{abstract}
\maketitle
\newpage
\twocolumngrid
\section{Introduction} Entanglement between different degrees of freedom (DoF) of a single particle is defined as single-particle entanglement (SPE). SPE furnishes further advantages for quantum-information-processing tasks over bipartite or multi-particle entanglement. The latter is more vulnerable to decoherence and is experimentally more resource-consuming than SPE~\cite{aqs,fang,gratsea_lewenstein_dauphin_2020}. Applications of SPE can be found in quantum networking via quantum joining~\cite{qjoining}, photonic quantum-information-processing, super dense coding~\cite{sdensecoding} and quantum key distribution~\cite{aqs,qkd1}. Further, SPE can be used in the analysis of photonic states and elementary particles~\cite{aqs}.

A single photon can exhibit entanglement (SPE) between any of the following three DoF, i.e., its orbital angular momentum (OAM), path and its polarization. The SPE between two nonlocal DoF of such a single particle, like the photon's internal OAM and its external path (i.e., the spatial trajectories), is an instance of a two-way SPE (or, nonlocal 2-way SPE)~\cite{Chandra2022}. Such nonlocal 2-way SPE states are crucial for encoding extensive quantum-information (QI) robustly, which will also be more resilient to decoherence~\cite{aqs}.

Moreover, a quantum state of a photon involving all of its three DoF, which is not separable with respect to any bi-partition in DoF, is said to possess genuine multipartite DoF entanglement or 3-way entanglement (SPE). This genuine multipartite DoF entanglement (GMDE) is equivalent to genuine multipartite entanglement (GME)~\cite{gme1,gme2,gme3}; in the latter, each subsystem refers to an individual particle state while in the former to an individual DoF of a single particle. The three-way SPE offers a significant advantage in quantum technology over two-way or bipartite entanglement. In particular, it has applications in large-scale quantum information~\cite{gmeuse} processing, state exploration and interaction among fundamental particles~\cite{gmeuse20,gmeuse2}. GME in multi-particle systems has been widely reported ~\cite{gme4,gme5,gme3,gme6}, and in Ref.~\cite{3wayGHZ} 3-way qubit-qubit-qubit SPE generation with photons (where each DoF is 2-dimensional) and its storage using a broadband solid-state-medium is reported. Note that, photonic 3-way SPE finds applications in designing hybrid quantum networks and repeaters that are useful for long-distance secure quantum communication~\cite{3wayGHZ}. However, there has been no attempt till date to generate and quantify genuine 3-way quNit-quNit-qubit SPE wherein two DoF are higher dimensional ($N>2$) while the third DoF is a qubit using any method, including quantum-walk setups or optical setups. This work bridges this crucial gap. The inclusion of higher dimensional DoF in 3-way (quNit-quNit-qubit) SPE, is crucial for encoding more information efficiently, as the quNit-DoF of the single-particle (like a photon) can increase communication channel capacity plus enhance quantum network capability~\cite{qudit1,qudit2,gmeuse,qjoining}.

A particle, like photon, can be mapped into a quantum walker in two spatial dimensions. The single photon's 3-way or nonlocal 2-way entanglement can be designed and controlled via the dynamics of a 2D discrete-time quantum walk (DTQW). In a 2D DTQW (quantum walk or QW), the walker moves in a 2D space, depending upon the coin operations. One can categorize 2D DTQWs based on the coin operations, see~\cite{BuschPRL,BuschPRA,Chandra2022}, as follows: 
\begin{itemize}
    \item[A.] Regular 2D QW: This uses a 4D coin, and the walker moves on $x$ and $y$-directions simultaneously~\cite{BuschPRL},
    \item[B.] Alternate 2D QW: This uses a 2D coin, and the walker moves on $x$ and $y$-directions alternately~\cite{BuschPRA}. 
\end{itemize}

Regular QWs, which use 4D coin operator, are experimentally very challenging~\cite{BuschPRL}. The 4D coin is a four-level
quantum system or two distinct qubits, which requires an entangling gate at every time-step which complicates the experimental-realization and only a few feasible physical implementations have been proposed so far like Ref.~\cite{4dcoin-expt}. The use of 2D coin reduces the experimental
challenges since it is just a single qubit~\cite{BuschPRL} and hence, alternate quantum walks(AQW) are resource-saving and have simpler physical realization. 

While SPE generation between two DoF, i.e., coin and position ($x$) of the walker via 1D DTQW has been explored extensively, see ~\cite{me-cb1, me-cb2,gratsea_lewenstein_dauphin_2020,fang,li_yan_he_wang_2018,ch2012disorder,vieira_amorim_rigolin_2013,vieira_amorim_rigolin_2014,gratsea2020universal,r_zhang2022}, there exists till date no attempt to generate genuine 3-way quNit-quNit-qubit entanglement (or, 3-way SPE) via a single particle evolution in 2D quantum walk setups. This 3-way SPE refers to the non-vanishing quantum correlations between all three DoF: $x$-position (quNit), $y$-position (quNit) and coin (qubit), of the single particle. The benefits of 3-way SPE over coin-position or local 2-way SPE motivates us to explore the applicability of resource-saving 2D AQW to design these 3-way entangled states.  Quantum speedup over classical computation and measurement-based quantum computing often necessitate a sufficiently large amount of entanglement~\cite{largespeprl}. In this paper, we also make an effort to maximise the generated 3-way SPE (quantified by $\pi$-tangle~\cite{fan,pitang}) to the highest possible value akin to symmetric W or GHZ-type tripartite states~\cite{gmeuse20,ghz1998}. Note that tracing over one DoF in a GHZ-type state results in a separable state, while doing the same for a W-type state results in a state with finite entanglement~\cite{gmeuse20}. In this context, we conjecture that the generated 3-way SPE states in this paper possess characteristics indicative of quNit-quNit-qubit W-type states.

Furthermore, a few studies on 2D regular and AQWs have been reported that generate 2-way (quNit-quNit) SPE between the walker's nonlocal $x$- and $y$-position DoF, see~\cite{BuschPRL,BuschPRA,Chandra2022}.The advantage 2-way (or 3-way) SPE has over bipartite or coin-position entanglement arises from the higher dimensionality of the position Hilbert spaces and exploitation of both the nonlocal ($x,y$) DoF of the single particle~\cite{aqs}. 
The Refs.~\cite{BuschPRL,BuschPRA,Chandra2022} discuss the nonlocal 2-way SPE generation with either a specific initial state~\cite{BuschPRL}  or fixed coin operator(s)~\cite{Chandra2022,BuschPRA}.
But a QW dynamics is highly dependent on both the choices of the initial state and the coin parameters and so is the SPE generation via the QW. Herein, in Sec.~\ref{s2}, we propose a generalized 2D AQW encompassing the most general separable initial state and an arbitrary single-qubit coin operator to generate nonlocal 2-way SPE as well as the genuine 3-way SPE states. The entanglement quantifiers, $\pi$-tangle for 3-way entanglement (between the three degrees of freedom) and negativity for nonlocal
2-way entanglement (between the nonlocal position degrees of freedom of the particle), are also detailed. We study the propensity of AQWs to generate the optimal amount of nonlocal 2-way and genuine 3-way SPE, from arbitrary initial separable quantum states, see Sec.~\ref{s3} and Sec.~\ref{s4}. We also observe one can control the amount of 3-way SPE or nonlocal 2-way SPE generated, by tuning the initial state and coin parameters. We finally discuss an experimental realization of this proposal using a single-photon circuit with passive optics, see Sec.~\ref{s5}. Sec.~\ref{s6} provides a conclusion on our results with some outlook. Following this in Appendices A-E, we deal with the derivation of CKW type inequalities for 2D AQW, we  establish the $\pi-$tangle as  a valid 3-way entanglement measure which fullfils all conditions for a  true measure of genuine multipartite (or, genuine multi DoF) entanglement, discussion on parameters which lead to maximal 3-way and maximal 2-way (nonlocal) GMDE, a comparison with other relevant schemes to generate non-local 2-way entanglement and finally a PYTHON code to generate some of the results.

\section{Generalized 2D AQW dynamics and generating single-particle entanglement}
\label{s2}
\begin{figure}[h!]
\includegraphics[width =9.1cm,height=2.65cm]{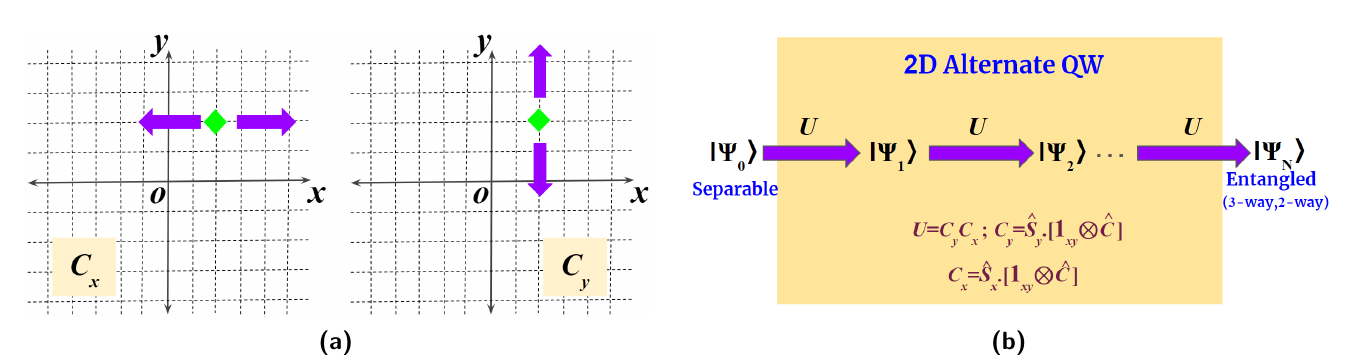}
\caption{(a) 2D AQW describing a quantum particle (green diamond) shifting either along $x$ or $y$-direction after a coin operation via $C_x$ or $C_y$ operator. (b) 2D AQW protocol for generating both genuine 3-way and nonlocal 2-way entanglement for a single-particle state, starting from an arbitrary separable initial state given in Eq.~(\ref{eq1}).}
\label{f1}
\end{figure}

AQW on 2D position space is defined as a tensor product $H_{xyc}=H_x\otimes H_y\otimes H_c$, where $H_x$ and $H_y$ are the infinite-dimensional $x$ and $y$ position Hilbert spaces, while $H_c$ is the two-dimensional coin Hilbert space in the computational basis $\{\ket{0_c},\ket{1_c}\}$. This can model a photon as the quantum walker. The photon's polarization states (e.g., horizontal or vertical $\{\ket{H},\ket{V}\}$) are mapped to the coin space $H_c$, while its path DoF corresponds to  $H_x$ and its OAM DoF aligns with $H_y$~\cite{Chandra2022}. For the quantum walker being initially localized at position-site $\ket{0_x,0_y}$ and with an arbitrary superposition over the coin states, the general initial state is then represented as $\ket{\psi_i}=\ket{\psi(t=0)}$, i.e.,
\begin{equation}
\ket{\psi_i} = \ket{0_x,0_y}\otimes[\cos(\frac{\theta}{2})\ket{0_c} + e^{i\phi}\sin(\frac{\theta}{2})\ket{1_c}],
\label{eq1}
\end{equation}
where $\theta \in [0, \pi]$ and phase $\phi \in [0, 2\pi)$.
The 2D AQW evolves by operating single-qubit coin operator $\hat{C}$ and the shift operators $\hat{S_x}$ and $\hat{S_y}$ respectively for the walker's movement in $x$ and $y$ directions (alternately).   
A general single-qubit coin operator is defined as,
\begin{equation}
\hat{C}(\alpha,\beta,\gamma) =
\begin{pmatrix}
\cos{\alpha} & e^{i\beta} \sin{\alpha}\\
\sin{\alpha}e^{i\gamma} & -e^{i(\beta+\gamma)} \cos{\alpha}
\end{pmatrix},
\label{eq2}
\end{equation}
where, $0\le\alpha,\beta,\gamma\le2\pi$.
The shift operators which move the walker along $x$ and $y$ directions are,
\begin{equation}
\begin{aligned}
\hat{S_x} = & \sum_{j=-\infty}^{\infty}\{\ket{(j-1)_x}\bra{j_x}\otimes\mathbb{1}_y\otimes\ket{0_c}\bra{0_c}\\
&+\ket{(j+1)_x}\bra{j_x}\otimes\mathbb{1}_y\otimes\ket{1_c}\bra{1_c}\},
\\
\hat{S_y} = & \sum_{j=-\infty}^{\infty}\{\mathbb{1}_x\otimes\ket{(j-1)_y}\bra{j_y}\otimes\ket{0_c}\bra{0_c}\\
&+\mathbb{1}_x\otimes\ket{(j+1)_y}\bra{j_y}\otimes\ket{1_c}\bra{1_c}\},
\end{aligned}
\label{eq3}
\end{equation}
where $\mathbb{1}_x$ and $\mathbb{1}_y$ are identity operators respectively in $x$ and $y$ Hilbert-spaces.
We model a generalized  2D AQW, wherein the full evolution operator is,
\begin{equation}
\begin{split}
U(t) = \hat{S}_y.[\mathbb{1}_{xy}\otimes \hat{C}].\hat{S}_x
.[\mathbb{1}_{xy}\otimes \hat{C}]\;,
\label{eq4}
\end{split}
\end{equation}
where $\hat{C}=\hat{C}(\alpha(t),\beta(t),\gamma(t))$ and $\mathbb{1}_{xy}=\mathbb{1}_{x}\otimes\mathbb{1}_{y}$. In line with Eq.~(\ref{eq4}), each time-step of the 2D AQW comprises the sequence: coin operation followed by the shift on x-axis, then coin operation followed by the shift on y-axis, as shown in Fig.~\ref{f1}(a). Equivalently, for each time-step, $U=C_xC_y$, with $C_x=\hat{S}_x
.[\mathbb{1}_{xy}\otimes \hat{C}]$ and $C_y=\hat{S}_y.[\mathbb{1}_{xy}\otimes \hat{C}]$. We call such an evolution sequence ($C_xC_yC_xC_y...$) a spatial sequence since the same coin is being applied to determine the particle's motion along both $x$- and $y$-directions, irrespective of the time step. 

It is important to note that at an arbitrary time step $t=n$, the quantum state of the particle (the walker) has the following general form:  
\begin{equation}
\ket{\psi_{n}} = U^n\ket{\psi_i}= \sum_{i,j}[h_{i,j}^{(n)}\ket{i_x,j_y,0_c}+v_{i,j}^{(n)}\ket{i_x,j_y,1_c}],
\label{eq5}
\end{equation}
where, $h_{i,j}^{(n)}$ and $v_{i,j}^{(n)}$ are respectively the normalized complex amplitudes of the states $\ket{i_x,j_y,0_c}$ and $\ket{i_x,j_y,1_c}$. These are functions of the intial state parameters $\{\theta,\phi\}$ and coin operator parameters $\{\alpha,\beta,\gamma\}$. On applying the coin (Eq.~(\ref{eq2})), the  particle can shift to either $x$ or $y$ direction. Consequently, the particle's state at the next time-step ($t=n+1$) will be,
\begin{align}
\begin{split} 
\ket{\psi_{n+1}} = &\sum_{i,j}[h_{i-1,j-1}^{(n+1)}\ket{(i-1)_x,(j-1)_y,0_c}\\& +h_{i+1,j-1}^{(n+1)}\ket{(i+1)_x,(j-1)_y,0_c}\\&+v_{i-1,j+1}^{(n+1)}\ket{(i-1)_x,(j+1)_y,1_c}\\&+v_{i+1,j+1}^{(n+1)}\ket{(i+1)_x,(j+1)_y,1_c}],
\end{split}
\label{eq6}
\end{align}

where, $h_{i-1,j-1}^{(n+1)}=\cos{\alpha}(h_{i,j}^{(n)}\cos{\alpha}+v_{i,j}^{(n)}e^{i\beta}\sin{\alpha})$,
$h_{i+1,j-1}^{(n+1)}=e^{i\beta}\sin{\alpha}(h_{i,j}^{(n)}e^{i\gamma}\sin{\alpha}-v_{i,j}^{(n)}e^{i(\beta+\gamma)}\cos{\alpha}),\;v_{i-1,j+1}^{(n+1)}=e^{i\gamma}\sin{\alpha}(h_{i,j}^{(n)}\cos{\alpha}+v_{i,j}^{(n)}e^{i\beta}\sin{\alpha}),\;v_{i+1,j+1}^{(n+1)}=e^{i(\beta+\gamma)}\cos{\alpha}(v_{i,j}^{(n)}e^{i(\beta+\gamma)}\cos{\alpha}-h_{i,j}^{(n)}e^{i\gamma}\sin{\alpha})$.

The interlacing coefficients within the time evolved states shown in Eqs.~(\ref{eq5}-\ref{eq6}) provide an indication that these states can exhibit SPE between the two nonlocal DoF: $x$-position (quNit) and $y$-position (quNit), which we identify as nonlocal 2-way (quNit-quNit) SPE, with $N=2n+1$ at $t=n$. We also see that we can have 3-way (quNit-quNit-qubit) SPE between coin (qubit), $x$-DoF and $y$-DoF, see Fig.~\ref{f1}(b).

We perform several simulations to find a wide range of spatial sequences comprising an arbitrary coin operator, which can generate nonlocal $xy$-SPE as well as 3-way SPE between the three DoF ($x,y$ and coin) from arbitrary separable initial states~(Eq.~(\ref{eq1})). 

\textcolor{brown}{\textit{Measuring entanglement.--}} 
The amount of quantum correlations in the single-particle entangled state between its three DoF ($x,y$ and coin), i.e., the 3-way SPE as well as between its nonlocal DoF ($x,y$), i.e., the nonlocal 2-way SPE can be measured via the negativity monotone~\cite{normneg,pitang, fan}. In this context, we first deal with generating 3-way  and then the nonlocal 2-way SPE in a single particle state evolving via the 2D AQW. The 3-way SPE of the quantum particle can be quantified by the monotone: triple-$\pi$ or $\pi$-tangle~\cite{fan,pitang}, i.e., $\pi_{xyc}$ and it is defined as $\pi_{xyc}=\frac{\pi_x+\pi_y+\pi_c}{3}$. Here, $\pi_i=N_{i|jk}^2- (N_{ij}^2+N_{ik}^2)$ where $N_{i|jk}=||\rho^{\text{T}_i}_{ijk}||-1$, $N_{ij}=||\rho^{\text{T}_i}_{ij}||-1$ with $i,j,k \in \{x,y,c\}$, and $\text{T}_i$ denotes partial transpose with respect to DoF $i$ while $||.||$ refers to matrix trace norm. The residual negativity values ($\pi_i$) satisfy the Coffman-Kundu-Wootters (CKW) type monogamy inequalities~\cite{fan,ckw},
$N_{i|jk}^2\ge N_{ij}^2+N_{ik}^2$, i.e., $\pi_i=N_{i|jk}^2- (N_{ij}^2+N_{ik}^2)\ge0$. Examples of QW states, Eq.~(\ref{eq6}), obeying the CKW-type inequalities, are given in Appendix~A. This is for the first time CKW-inequalities have been shown to be obeyed in a high-dimensional system. It was shown to work only for qubit-qubit-qubit system. Further, we prove that the $\pi$-tangle satisfies all conditions necessary for a faithful GME or GMDE measure~\cite{gmem1,gme1,lu1,neg2002,locc}, see Appendix~B.
The average of the $\pi$-tangle at fixed $\phi$ (where $\phi$ is a parameter in initial state Eq.~(\ref{eq1})) is, $\pi_{av}=\langle \pi_{xyc} \rangle = \frac{1}{\pi} \int_{0}^{\pi} \;\pi_{xyc}\;d\theta$. In general, $\pi_{xyc},\pi_{av}\ge0$ whereas $\pi_{xyc}=\pi_{av}=0$  for any biseparation (e.g., for the initial state Eq.~(\ref{eq1})), i.e., in the absence of genuine 3-way  entanglement~\cite{fan}. 

To quantify the degree of entanglement between the nonlocal $x$ and $y$ DoF of the particle, we use entanglement negativity ($N$)~\cite{neg2002}. To compute $N$ at any QW-time-step $t$, we first take the partial trace of the time-evolved density operator $\rho_{xyc}=\ket{\psi(t)}\bra{\psi(t)}$ over the coin DoF, i.e., $\rho_{xy}=\text{Tr}_c(\rho_{xyc})$, which is a quantum mixed state. From the eigenvalues $\{\lambda_i: i\in[1,(2t+1)^2]\}$ of the partial transpose (with respect to $x$ or $y$ DoF) of the reduced density matrix $\rho_{xy}$, we find the entanglement negativity,
$N=\sum_{i=1}^{(2t+1)^2}\frac{|\lambda_i|-\lambda_i}{2}$. 
Positive $N$ values refer to entangled states, and for any separable or non-entangled state like the initial state given by Eq.~(\ref{eq1}), $N=0$. With a fixed $\phi$, average negativity is, $N_{av}=\langle N \rangle = \frac{1}{\pi} \int_{0}^{\pi}\;N\;d\theta$, and clearly, for a separable initial state (say, Eq.~(\ref{eq1})), $N_{av}=0$.

\begin{figure}[h!]
\includegraphics[width=9cm,height=4.3cm]{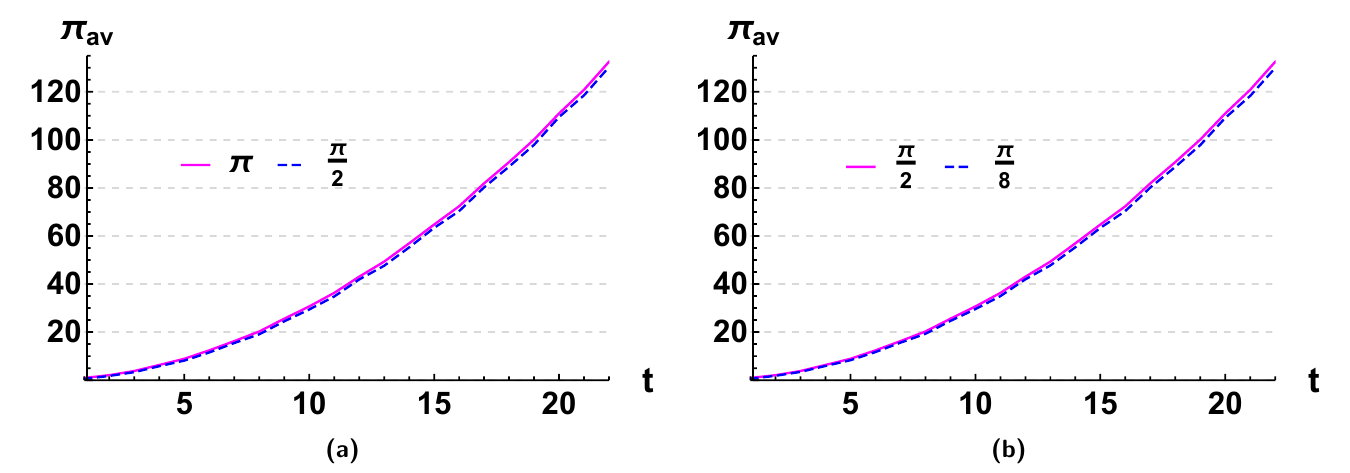}
\caption{Three-way entanglement  $\pi_{av}$ vs time steps($t$) for spatial evolution sequences: (a) $M1_xM1_y...$ for arbitrary separable initial state (Eq.~(\ref{eq1})) with $\phi=\pi,\frac{\pi}{2}$; (b) $M2_xM2_y...$ for an arbitrary separable initial state with $\phi=\frac{\pi}{8},\frac{\pi}{2}$. Note that at all time steps, $M1_xM1_y$ with $\phi=\pi$  and $M2_xM2_y$ with $\phi=\frac{\pi}{2}$ yield the best values for 3-way entanglement among the cases we have studied.}
\label{f2}
\end{figure}

\section{3-way entanglement generation}
\label{s3}
A quantum state of the particle (quantum walker) having 3-way SPE (i.e., nonzero $\pi$-tangle) is not separable with respect to any bi-partition of the DoF.  One can design quantum coin operators corresponding to different $\phi$ (phase parameter of the initial state) values that generate the maximum possible 3-way entanglement for the 2D AQW. Two such evolution operators (see, Eq.~\ref{eq4}) are, $M1_xM1_y,$ with $\phi=0,\pi,2\pi$ and $M2_xM2_y$ with $\phi=\frac{\pi}{2},\frac{3\pi}{2}$. Coin $\hat{M1}=\hat{C}(\alpha=\frac{5\pi}{16},\beta=\frac{\pi}{2},\gamma=\frac{\pi}{2})$, while coin $\hat{M2}=\hat{C}(\alpha=\frac{5\pi}{16},\beta=\pi,\gamma=\frac{\pi}{4})$. For evolution up to time step $t=22$, Fig.~\ref{f2}(a) shows the generated 3-way SPE for evolution sequences, $M1_xM1_y...$ with initially separable states Eq.~(\ref{eq1}) and $\phi=\pi,\frac{\pi}{2}$, while Fig.~\ref{f2}(b) depicts the 3-way SPE for evolution sequence $M2_xM2_y...$ with the same initial state (Eq.~(\ref{eq1})) with $\phi=\frac{\pi}{2},\frac{\pi}{8}$. We clearly see from Fig.~\ref{f2} that $M1_xM1_y$ gives maximal average $\pi$-tangle for $\phi=\pi$, whereas $M2_xM2_y$ gives the best results for $\phi=\frac{\pi}{2}$. Details regarding the simulations and the coins used to achieve maximal 3-way entanglement ($\pi_{av}$) in 2D AQW as a function of initial state parameters are provided in Appendix~C. The above-mentioned coins ($M1,\;M2$) are two of the numerous optimal coins (best entangling coins obtained from the simulations) that can generate maximal 3-way SPE in 2D QW setups; see Appendix~C and Table~I.
The evolution $M1_xM1_y$ with the initial state parameter $\phi=\pi$ yields $\pi_{av}=2.0656 \;(\text{at } t=2),30.6639\;(\text{at }t=10),132.5407\;(\text{at }t=22)$, whereas $M2_xM2_y$ with the state parameter $\phi=\frac{\pi}{2}$ yields  $\pi_{av}=2.0656 \;(\text{at }t=2),30.6639\;(\text{at }t=10), 132.5407\;(\text{at }t=22)$, see Fig.~\ref{f2}. 

\begin{figure}[h]
\includegraphics[width=9.3cm,height=8.5cm]{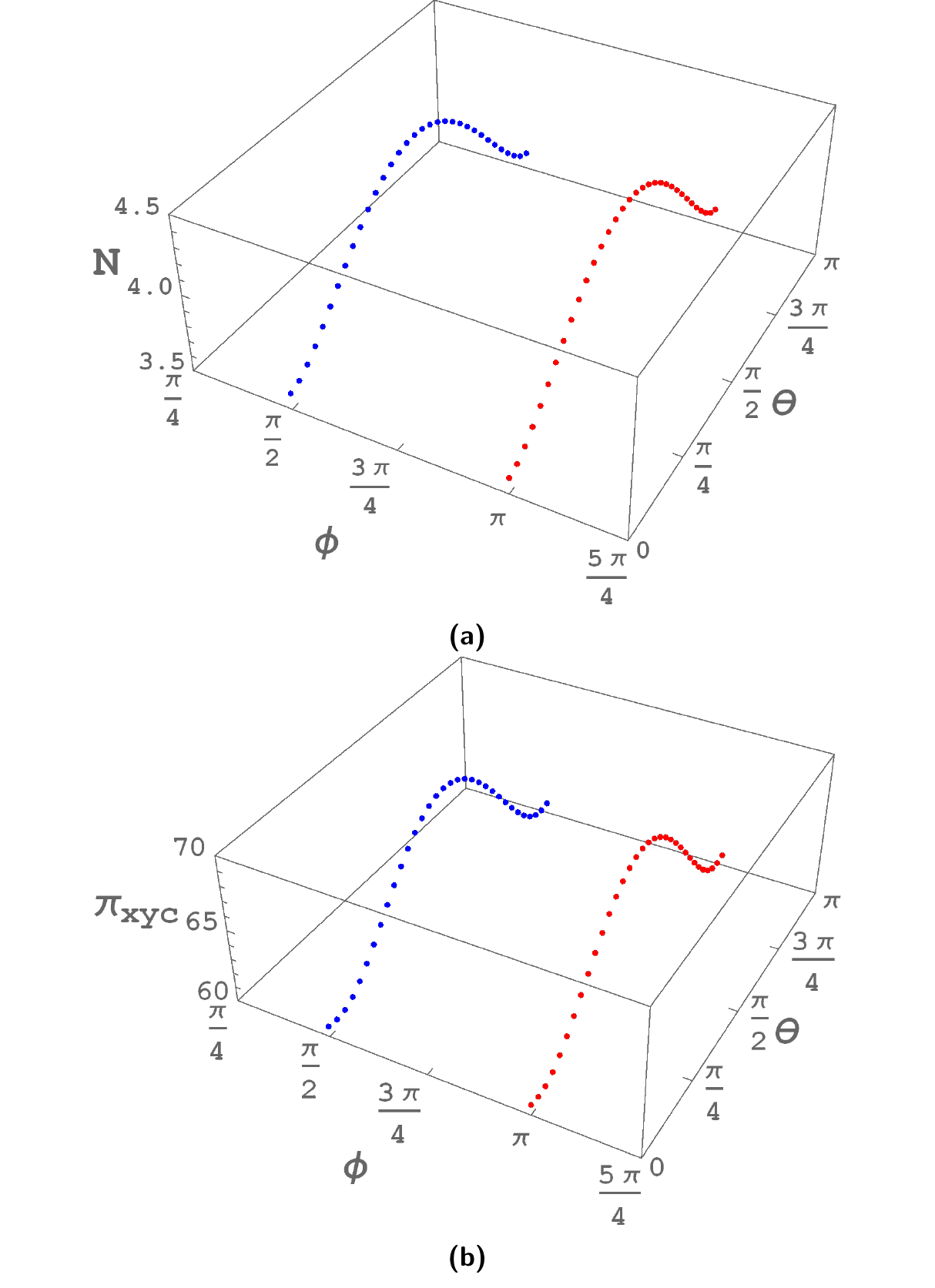}
\caption{(a) 3-way entanglement ($\pi_{xyc}$) between $x,y$ and coin DoF vs the initial state parameter $\theta$ for spatial evolution sequences: $M1_xM1_y$ for the  phase variable (of the initial state) $\phi=\pi$,  and $M2_xM2_y$ for $\phi=\frac{\pi}{2}$ at time step $t=15$; (b) Nonlocal 2-way entanglement ($N$) between $x,y$ DoF vs the initial state parameter $\theta$ for spatial evolution sequences: $G1_xG1_y$ for the  phase variable (of the initial state) $\phi=\pi$, $G2_xG2_y$ for $\phi=\frac{\pi}{2}$ at time step $t=15$. }
\label{f3}
\end{figure}

Further, in Fig.~\ref{f3}(a), we show the amount of 3-way SPE $\pi_{xyc}$ generated by both the evolution operators: $M1_xM1_y,\;M2_xM2_y$ (with corresponding optimal $\phi$ values), at $t=15$ as a function of initial state parameter $\theta$ (which goes from 0 to $\pi$ in steps of $\frac{\pi}{32}$). This $\theta$-tuning reveals that for $\phi=\pi$ and $\theta=\frac{\pi}{2}$, the sequence $M1_xM1_y...$ yields maximum possible 3-way SPE with $\pi_{xyc}=69.0024$ (at $t=15$). At $\phi=\frac{\pi}{2}$ and $\theta=\frac{\pi}{2}$, the sequence $M2_xM2_y...$ yields maximum possible 3-way SPE with $\pi_{xyc}=69.0024$ again. For more analysis on maximal 3-way SPE generating evolution sequences, see Appendix~C. 

\section{Nonlocal 2-way $xy$-entanglement generation}
\label{s4}
One can form numerous coin operators for arbitrary initial state $\phi$ values to generate maximum possible entanglement (SPE) between the nonlocal ($x,y$) DoF of the particle that evolves via the 2D AQW. Two distinct coin operators from Eq.~(\ref{eq2}), which generate maximum possible nonlocal $xy$ SPE are, $G1=\hat{C}(\alpha=\frac{19\pi}{16},\beta=\frac{\pi}{2},\gamma=\frac{\pi}{2})$ with $\phi=0,\pi$ and $G2=\hat{C}(\alpha=\frac{19\pi}{16},\beta=\pi,\gamma=\frac{\pi}{16})$ with $\phi=\frac{\pi}{2},\frac{3\pi}{2}$. See Appendix~C for details on our simulations to find such coins. We denote the evolution operators involving these coins (Eq.~(\ref{eq4})) to be $G1_xG1_y$ and $G2_xG2_y$.

\begin{figure}[h]
\includegraphics[width =9cm,height=4.3cm]{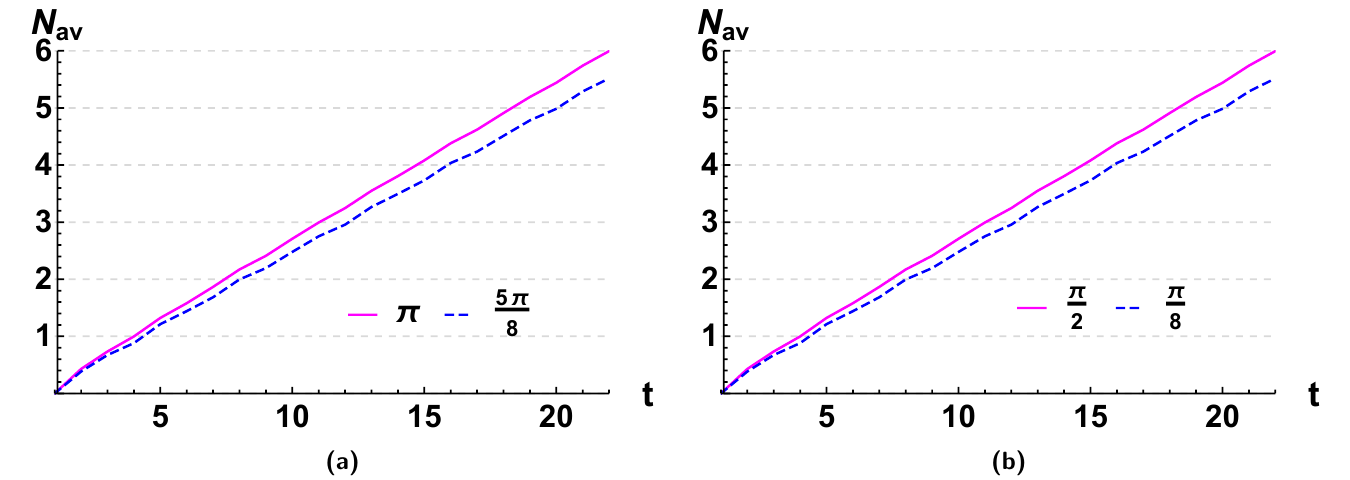}
\caption{Average nonlocal 2-way entanglement ($N_{av}$) between $x,y$ DoF  vs time steps($t$) for spatial evolution sequences: (a) $G1_xG1_y...$ for the arbitrary separable initial state with $\phi=\pi,\frac{5\pi}{8}$, (b) $G2_xG2_y...$ for arbitrary separable initial state with $\phi=\frac{\pi}{8},\frac{\pi}{2}$. Note that $G1_xG1_y$ and $G2_xG2_y$ yield maximum 2-way entanglement respectively for $\phi=\pi$ and $\phi=\frac{\pi}{2}$.}
\label{f4}
\end{figure}

Fig.~\ref{f4} shows the average negativity ($N_{av}$) generated via the evolution  $G1_xG1_y$ with $\phi=\pi,\frac{5\pi}{8}$ and $G2_xG2_y$ with $\phi=\frac{\pi}{8},\frac{\pi}{2}$. The spatial evolution sequences $G1_xG1_y...\;, G2_xG2_y...$ yield maximal average entanglement negativity respectively for $\phi=\pi$ and $\phi=\frac{\pi}{2}$, see Fig.~\ref{f4}, as one expects from the simulation (see, Appendix~C). Specifically, at time steps $t=2,10,22$, the sequence $G1_xG1_y$ for $\phi=\pi$ yields corresponding $N_{av}=0.4290,2.7089,5.9950$, while the sequence $G2_xG2_y$ for $\phi=\frac{\pi}{2}$ yields almost the same values as $G1_xG1_y$, see also Appendix~C for more such optimal coins.  

Furthermore, in Fig.~\ref{f3}(b), we show the amount of nonlocal 2-way entanglement $N$ generated by both the spatial evolution sequences at $t=15$ as a function of initial state parameter $\theta$ (which goes from 0 to $\pi$ in steps of $\frac{\pi}{32}$). This reveals that at $\phi=\pi$ and $\theta=\frac{\pi}{2}$, the sequence $G1_xG1_y...$  yields maximum nonlocal 2-way ($xy$) SPE with $N=4.4429$. Similarly, at $\phi=\frac{\pi}{2}$ and $\theta=\frac{\pi}{2}$, the spatial sequence $G2_xG2_y...$  also yields the same maximum nonlocal 2-way ($xy$) SPE with $N=4.4429$.

\textcolor{brown}{\textit{Remarks.--}}
The above numerical simulations and analysis with the most general initial state (Eq.~\ref{eq1}) yield numerous coin operators corresponding to certain phase values ($\phi$), which can generate a maximum amount of SPE between all three ($x,y$ and coin) or the nonlocal ($x,y$) DoF of the single quantum particle evolving via the 2D AQW dynamics. For more coin evolution sequences like $M1_xM1_y,\;M2_xM2_y$ for 3-way SPE and $G1_xG1_y,\;G2_xG2_y$ for 2-way SPE and the numerical optimization details, see Appendix~C. The scheme also predicts at what specific value of state parameters $\phi$ and $\theta$, one can obtain maximum entanglement with the optimal coin evolution sequences. This contributes towards a state-of-art optimum entanglement generating protocol via 2D AQW. In addition, by tuning the parameters $\phi$ and $\theta$ of the initial state, one obtains controlled entanglement dynamics via the 2D AQW setup. 

\section{Photonic 2D AQW setup to generate 3-way \& nonlocal 2-way single-particle entanglement}
\label{s5}
{Herein, we demonstrate via a single-photon circuit with passive optical devices such as Jones-plates (J-plates) and polarizing beam splitters (PBS) that our proposal can be experimentally realized with a single-photon~\cite{Chandra2022,expt1-science,expt2,expt3,expt4,expt5,2dqw-expt,jplate} wherein the photon's OAM, path and polarization are mapped respectively into the walker's $x,\;y$ and coin DoF.} 
\begin{figure}[h]
\includegraphics[width =8cm,height=4.65cm]{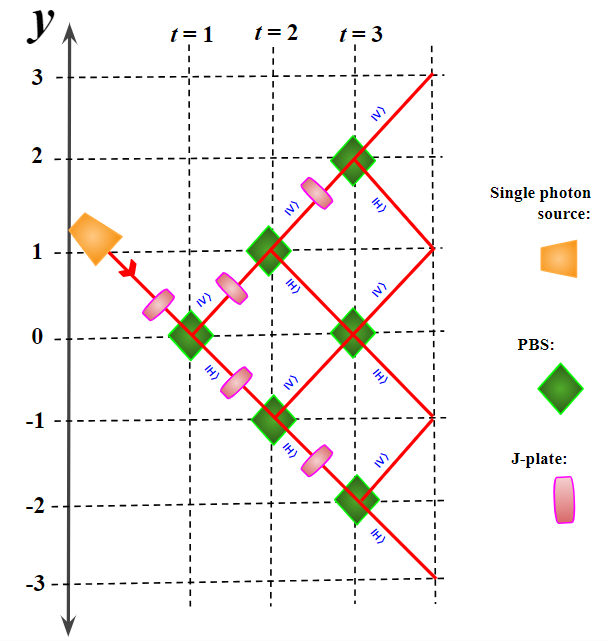}
\caption{Single-photon based realization for generating genuine 3-way and nonlocal 2-way single-particle entangled states via 2D AQW at time steps up to $t=3$.}
\label{fs7}
\end{figure}{A PBS can act as operator $\hat{S}_y$ in the 2D AQW dynamics, as a PBS transmits vertical polarization  ($\ket{1_c}$) while reflecting horizontal polarization ($\ket{0_c}$). On the other hand, depending on the polarization state of a photon, a J-plate can modify the photon's OAM, which enables this device to emulate the AQW evolution operation: qubit-coin $\rightarrow$ a shift in $x$ axis $\rightarrow$ qubit-coin. Thus, a J-plate operation is equivalent to, $J(\alpha,\beta,\gamma)\equiv[\mathbb{1}_{xy}\otimes \hat{M1}]).\hat{S}_x
.[\mathbb{1}_{xy}\otimes \hat{M1}]$ to generate optimal genuine 3-way SPE, or $J(\alpha,\beta,\gamma)\equiv[\mathbb{1}_{xy}\otimes \hat{G1}]).\hat{S}_x
.[\mathbb{1}_{xy}\otimes \hat{G1}]$ to generate optimal nonlocal 2-way SPE, also see Figs.~\ref{f2} and \ref{f4}. Fig.~\ref{fs7} depicts a schematic representation of the photonic implementation of our AQW-scheme to generate the genuine 3-way and nonlocal 2-way single-photon entangled states. The output photon from this setup, has 3-way and/or nonlocal 2-way SPE between its OAM, path and polarization. The SPE quantifiers, $\pi$-tangle and entanglement negativity, can be computed via a full-state tomography or an ML-assisted measurement method using a swap-operator series
on several copies of the entangled photonic state~\cite{negexpt}.}

\section{Summary with outlook}
\label{s6}
Herein, we develop a framework to generate and control genuine 3-way (quNit-quNit-qubit) SPE involving all $x,y,c$ DoF, as well as nonlocal 2-way (quNit-quNit) SPE (involving $x$ and $y$ DoF) of any quantum particle evolving via a generalised 2D AQW with resource-saving single-qubit coins. The 3-way SPE and 2-way SPE are quantified by $\pi$-tangle (see Appendix~A for CKW-inequality in the AQW system and Appendix~B for the proof that $\pi$-tangle is a true GMDE/GME measure) and negativity respectively. The analytical-cum-numerical investigations predict numerous optimal quantum coin operators corresponding to particular values of the initial state ($\phi$), which yield maximal 3-way and 2-way nonlocal SPE for the particle evolving from an initial separable state via the QW setup (also see Appendix~C). The optimal coins for 3-way SPE also yield a large amount of 2-way SPE, which implies that the generated 3-way SPE states can be predominantly of W-type~\cite{gmeuse20}. A comparison of our work with other relevant works~\cite{BuschPRA,BuschPRL,Chandra2022} is given in Appendix~D. Our work, apart from opening an unprecedented avenue for generating 3-way quNit-quNit-qubit SPE, outperforms the other existing schemes in terms of overall resource-saving architecture and yielding high values of nonlocal 2-way quNit-quNit SPE too. We provide a PYTHON code to obtain the numerical results in Appendix~E.

{We demonstrate via a single-photon circuit with passive optical devices the implementation of our 3-way and 2-way entanglement-generating scheme.} Our scheme can also be adapted to other experimental setups, e.g., using trapped ions~\cite{ion1,ion2} or trapped and ultra-cold atoms~\cite{atom1,atom2,atom3}, that can realize 2D AQW. This scheme contributes towards state-of-art in generating genuine 3-way entangled and 2-way entangled single-particle states, which are crucial for large-scale QI processing. Our scheme of designing and controlling entanglement dynamics via quantum walks have potential applications in entanglement protocols for quantum information processing~\cite{aqs}, QW-based quantum algorithms~\cite{algo2D}, hybrid quantum networks for long-range quantum-secure communication~\cite{3wayGHZ}, realization of non-Markovian
quantum channels~\cite{chandra2}, studying the dynamics of open quantum
systems and entanglement-based quantum cryptography~\cite{crypt15,p4}.

\onecolumngrid


\section{Appendix}

In the Appendices A-E, we provide a more detailed derivation of the validity of the CKW-type monogamy inequalities~\cite{ckw,fan} for the 2D alternate quantum walk (AQW) states in Appendix~A. In Appendix~B, we provide the proof that $\pi$-tangle obeys all conditions for a true GMDE/GME measure for genuine 3-way single-particle entanglement (SPE). In Appendix~C, we give details of simulations to obtain coin and initial state parameters that yield a maximal amount of 3-way (quNit-quNit-qubit) and nonlocal 2-way (quNit-quNit) SPE. Further details on 3-way and 2-way SPE generation via the optimal coins designed via the simulations are included, also. Appendix~D provides a comparison of our work with other relevant proposals to generate nonlocal 2-way SPE. Finally, we provide a Python code to generate figures of our work in Appendix~E for interested researchers.  
\appendix

\section{ CKW-type inequalities in single-particle entanglement via 2D alternate quantum walk (AQW)}
\label{appa}

The 3-way entanglement of the quantum particle evolving via the 2D alternate QW (AQW) (as discussed in the main text) can be quantified by the monotone triple-$\pi$ or $\pi$-tangle~\cite{fan,pitang}, i.e., $\pi_{xyc}$. 
As in the main text (Eq.~(1)), the initial separable state of the particle (quantum walker) at time-step $t=0$ is $\ket{\psi_i}=\ket{\psi(t=0)}$, i.e.,
\begin{equation}
\ket{\psi_i} = \ket{0_x,0_y}\otimes[\cos(\frac{\theta}{2})\ket{0_c} + e^{i\phi}\sin(\frac{\theta}{2})\ket{1_c}],
\label{eq7}
\end{equation}

where $\theta \in [0, \pi]$ and phase $\phi \in [0, 2\pi)$. The particle evolves via the evolution operator,
\begin{equation}
\begin{split}
U(t) = \hat{S}_y.[\mathbb{1}_{xy}\otimes \hat{C}].\hat{S}_x
.[\mathbb{1}_{xy}\otimes \hat{C}]\;,
\label{eq8}
\end{split}
\end{equation}
where $\hat{C}=\hat{C}(\alpha(t),\beta(t),\gamma(t))=\begin{pmatrix}
\cos{\alpha} & e^{i\beta} \sin{\alpha}\\
\sin{\alpha}e^{i\gamma} & -e^{i(\beta+\gamma)} \cos{\alpha}
\end{pmatrix}$ is the coin operator and 
\begin{equation}
\begin{aligned}
\hat{S_x} = & \sum_{j=-\infty}^{\infty}\{\ket{(j-1)_x}\bra{j_x}\otimes\mathbb{1}_y\otimes\ket{0_c}\bra{0_c}+\ket{(j+1)_x}\bra{j_x}\otimes\mathbb{1}_y\otimes\ket{1_c}\bra{1_c}\},
\\
\hat{S_y} = & \sum_{j=-\infty}^{\infty}\{\mathbb{1}_x\otimes\ket{(j-1)_y}\bra{j_y}\otimes\ket{0_c}\bra{0_c}+\mathbb{1}_x\otimes\ket{(j+1)_y}\bra{j_y}\otimes\ket{1_c}\bra{1_c}\},
\end{aligned}
\label{eq9}
\end{equation}

are the shift operators that move the walker along $x$ and $y$ directions. $\mathbb{1}_x$ and $\mathbb{1}_y$ are identity operators respectively in $x$ and $y$ position Hilbert-spaces. The evolution operator (Eq.~\ref{eq8}) for each QW time-step is denoted by, $C_xC_y$, with $C_x=\hat{S}_x
.[\mathbb{1}_{xy}\otimes \hat{C}]$ and $C_y=\hat{S}_y.[\mathbb{1}_{xy}\otimes \hat{C}]$.

\begin{figure}[H]
\includegraphics[width = 18cm,height=6cm]{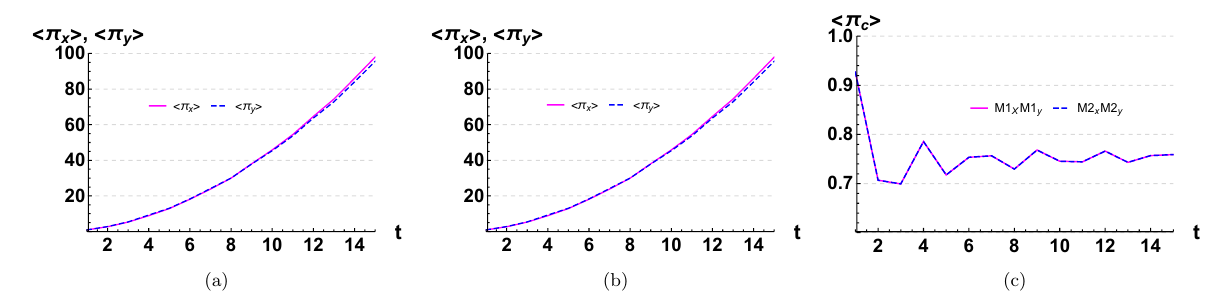}
\caption{CKW-type inequalities for quantum states in alternate 2D QWs: (a) Average residual negativities ($\langle \pi_{x}\rangle$, $\langle \pi_{y}\rangle$) vs time steps($t$) for spatial evolution sequence $M1_xM1_y...$ for an separable initial state Eq.~(\ref{eq7}) with $\phi=\pi$; (b) Average residual negativities ($\langle \pi_{x}\rangle$, $\langle \pi_{y}\rangle$) vs time steps($t$) for spatial evolution sequence $M2_xM2_y...$ for separable initial state Eq.~(\ref{eq7}) with $\phi=\frac{\pi}{2}$; (c) Average residual negativity ($\langle \pi_{c}\rangle$) vs time steps($t$) for spatial evolution sequences, $M1_xM1_y...$ for arbitrary separable initial state Eq.~(\ref{eq7}) with $\phi=\pi$ and $M2_xM2_y...$ for the separable initial state with $\phi=\frac{\pi}{2}$.}
\label{fs1}
\end{figure}

For the particle's states at any QW-time-step $t$, the following residual negativity values need to satisfy the Coffman-Kundu-Wootters (CKW) type monogamy inequalities~\cite{fan,ckw}: 
$N_{i|jk}^2\ge N_{ij}^2+N_{ik}^2$, where $N_{i|jk}=||\rho^{\text{T}_i}_{ijk}||-1$, $N_{ij}=||\rho^{\text{T}_i}_{ij}||-1$ with $i,j,k \in \{x,y,c\}$, and $\text{T}_i$ denotes partial transpose with respect to DoF $i$ while $||.||$ refers to matrix trace norm. This is equivalent to, 
\begin{equation}
\begin{aligned}
\pi_x=&N_{x|yc}^2- (N_{xy}^2+N_{xc}^2)\ge0,
\;
\pi_y=&N_{y|cx}^2- (N_{yc}^2+N_{yx}^2)\ge0,\;
\text{and }\pi_c=&N_{c|xy}^2- (N_{cx}^2+N_{cy}^2)\ge0,
\end{aligned}
\label{eq10}
\end{equation}

i.e., $\pi_x,\pi_y,\pi_c$ are non-negative. Thus, at a fixed $\phi$-value, the $\theta$-average values $\langle \pi_{i} \rangle = \frac{1}{\pi} \int_{0}^{\pi} \;\pi_{i}\;d\theta$ would follow the inequalities $\langle \pi_{i}\rangle \ge 0$, i.e., $\langle \pi_{i} \rangle$ is non-negative for all $i\in \{x,y,c\}$ at any $t$. For examples of quantum states generated via the alternate 2D quantum walk (QW) that satisfies the inequalities, see Fig.~\ref{fs1}. In  Fig.~\ref{fs1}, the single-particle states are generated via two QW evolution sequences: $M1_xM1_y...$  and $M2_xM2_y...$ with state phase parameter $\phi=\pi$ or $\phi=\frac{\pi}{2}$. Fig.~\ref{fs1}(a) and (b) respectively show the plots of $\langle \pi_{x} \rangle$, $\langle \pi_{y} \rangle$ for the  the sequences $M1_xM1_y...$ and $M2_xM2_y...$ , as a function of time step $t$. In Fig.~\ref{fs1}(c) we show the plot of $\langle \pi_{c} \rangle$ for the sequences $M1_xM1_y...$ and $M2_xM2_y...$ , as a function of time step $t$. Clearly, the values $\langle \pi_{i} \rangle$ are non-negative for all $i\in \{x,y,c\}$ at any time step $t$. This shows that QW states do satisfy the CKW-type inequalities.

The 3-way entanglement measure, $\pi$-tangle is defined as $\pi_{xyc}=\frac{\pi_x+\pi_y+\pi_c}{3}$ and its average at fixed $\phi$ is, $\pi_{av}=\langle \pi_{xyc} \rangle = \frac{1}{\pi} \int_{0}^{\pi} \;\pi_{xyc}\;d\theta$. The validity of CKW-type inequalities ($\pi_x,\pi_y,\pi_c\ge0$) implies that, in general, the 3-way entanglement measure $\pi_{xyc}\ge0$. Note that $\pi_{xyc}=\pi_{av}=0$ for any biseparation, i.e., in the absence of genuine 3-way  entanglement. For genuine 3-way  entanglement, the quantum correlations are shared between all three DoF ($x,y$, coin) of the quantum particle~\cite{fan}. The evolution sequences $M1_xM1_y...$ and $M2_xM2_y...$ generate genuine 3-way SPE ($\pi_{av}>0$) at all time-steps as shown in the main, see Fig.~2.

\section{ PROOF--The $\pi$-tangle obeys all  conditions for a true GME measure of high-dimensional 3-way entanglement} 
\label{appb}
A measure of genuine multipartite DoF entanglement (GMDE) or genuine multipartite entanglement (GME) must obey the following five conditions (C1-C5) to qualify as a faithful GMDE/GME measure~\cite{gmem1,gme1}. We note that a 3-way SPE is also a GME between 3-DoFs of a single-particle, as discussed in the main text introduction and we employ $\pi$-tangle to compute the amount of genuine 3-way entanglement between the 3 DoFs, i.e., quNit-quNit-qubit system. Herein we show, for the first time, how $\pi$-tangle obeys all of these five conditions to qualify as a faithful GME measure for our AQW setup.
\begin{itemize}
    \item[\textbf{C1}.] \underline{Statement}: The GME measure has to be 0 for separable and biseparable states.
    
    \underline{Proof}:
    The genuine 3-way entanglement measure, $\pi$-tangle is defined as,
\begin{equation}
\pi_{xyc}=\frac{\pi_x+\pi_y+\pi_c}{3} \text{ where }
\pi_x=N_{x|yc}^2- (N_{xy}^2+N_{xc}^2),
\;
\pi_y=N_{y|cx}^2- (N_{yc}^2+N_{yx}^2),\;
\pi_c=N_{c|xy}^2- (N_{cx}^2+N_{cy}^2).
\label{gme}
\end{equation}
As negativity ($N_{x|yc}, N_{xy}$ etc.) is zero for product states like, $\ket{0_x}\otimes\ket{0_y}\otimes\ket{0_c}$, from Eq.~(\ref{gme}), $\pi_{xyc}=0$. Furthermore, for biseparable states, such as $\ket{0_x}\otimes\frac{(\ket{0_y,0_c}+\ket{1_y,1_c})}{\sqrt{2}}$ ($x$-DoF is unetangled from the remaining two DoF), $\frac{(\ket{0_x,0_y,0_c}+\ket{1_x,0_y,1_c})}{\sqrt{2}}$ ($y$-DoF is unetangled from the remaining two DoF), $\frac{(\ket{0_x,0_y}+\ket{1_x,1_y})}{\sqrt{2}}\otimes\ket{0_c}$ ($c$-DoF is unetangled from the remaining two DoF), we calculate the $\pi$-tangle $\pi_{xyc}$.  For the biseprable state, $\ket{0_x}\otimes\frac{(\ket{0_y,0_c}+\ket{1_y,1_c})}{\sqrt{2}}$, we find that, $N_{x|yc}=0, N_{c|xy}=1,N_{y|cx}=1, N_{xy}=N_{yx}=0,N_{xc}=N_{cx}=0,N_{yc}=N_{cy}=1$ and thus, using Eq.~(\ref{gme}), we get $\pi_{xyc}=0$. Again for the biseparable state, $\frac{(\ket{0_x,0_y,0_c}+\ket{1_x,0_y,1_c})}{\sqrt{2}}$, we find, $N_{x|yc}=1, N_{c|xy}=1,N_{y|cx}=0, N_{xy}=N_{yx}=0,N_{xc}=N_{cx}=1,N_{yc}=N_{cy}=0$ and thus, using Eq.~(\ref{gme}), we obtain $\pi_{xyc}=0$. Finally, for the biseparable state, $\frac{(\ket{0_x,0_y}+\ket{1_x,1_y})}{\sqrt{2}}\otimes\ket{0_c}$, we estimate that, $N_{x|yc}=1, N_{c|xy}=0,N_{y|cx}=1, N_{xy}=N_{yx}=1,N_{xc}=N_{cx}=0,N_{yc}=N_{cy}=0$ and hence, using Eq.~(\ref{gme}), we find $\pi_{xyc}=0$.  Thus, $\pi$-tangle is zero for biseparable and product states, and thus it satisfies C1.  Such result is also supported by the fact that the residual negativities $\pi_x,\pi_y,\pi_c$ obey CKW inequality, i.e., $N_{i|jk}^2\ge N_{ij}^2+N_{ik}^2$, where $i,j,k \in \{x,y,c\}$, as discussed above in Appendix~A.
     \item[\textbf{C2}.] \underline{Statement}: The GME measure has to be positive for GME states or 3-way entangled single-particle states.
     
     \underline{Proof}:
     We observe that $\pi$-tangle is positive for GHZ state ($\pi_{xyc}=1$) and W state ($\pi_{xyc}\approx0.5493$), as also observed to be positive in Ref.~\cite{fan}, which implies $\pi$-tangle satisfies C2.   Fig.~\ref{fs1} above shows the average residual entanglements,  $\pi_x,\pi_y,\pi_c$ are positive, which in turn yields average $\pi$-tangle to be positive or indicates the existence of genuine 3-way SPE in our AQW setup.
     
     \item[\textbf{C3}.] \underline{Statement}: The GME measure has to be invariant under local unitary (LU) transformations.
     
     \underline{Proof}: The nine negativities, $N_{x|yc}, N_{xy}, N_{xc}, N_{y|cx}, N_{yc},N_{yx}, N_{c|xy}, N_{cx},N_{cy}$ are proven LU invariants ~\cite{fan, lu1,neg2002} and thus are their squared values, i.e.,  $N_{x|yc}^2, N_{xy}^2, N_{xc}^2, N_{y|cx}^2, N_{yc}^2,N_{yx}^2, N_{c|xy}^2, N_{cx}^2,N_{cy}^2$ and hence $\pi$-tangle is a LU invariant too. Thus $\pi$-tangle obeys C3.
     
     \item[\textbf{C4}.] \underline{Statement}: The GME measure has to be nonincreasing under LOCC (local operations
and classical communication).

\underline{Proof}:
Negativity does not increase under LOCC~\cite{locc}, so $\pi$-tangle is nonincreasing under LOCC, and hence, it satisfies C4.

\item[\textbf{C5}.] \underline{Statement}: The GME measure has to rank the well-known GHZ state to be more entangled as compared to the W-state  or flipped W-state.

\underline{Proof}:
We estimate that $\pi$-tangle for a GHZ state, i.e., $\frac{1}{\sqrt{2}}(\ket{0_x,0_y,0_c}+\ket{1_x,1_y,1_c})$ is 1 and that for a W-state, i.e., $\frac{1}{\sqrt{3}}(\ket{1_x,0_y,0_c}+\ket{0_x,1_y,0_c}++\ket{0_x,0_y,1_c})$  is 0.54936354 (approx.). Similarly for a flipped W-state, i.e., $\frac{1}{\sqrt{3}}(\ket{1_x,1_y,0_c}+\ket{1_x,0_y,1_c}++\ket{0_x,1_y,1_c})$, the $\pi$-tangle is again 0.54936354 (approx.). Therefore, $\pi$-tangle obeys C5. 

\end{itemize}

\begin{figure}[H]
\includegraphics[width = 18cm,height=13cm]{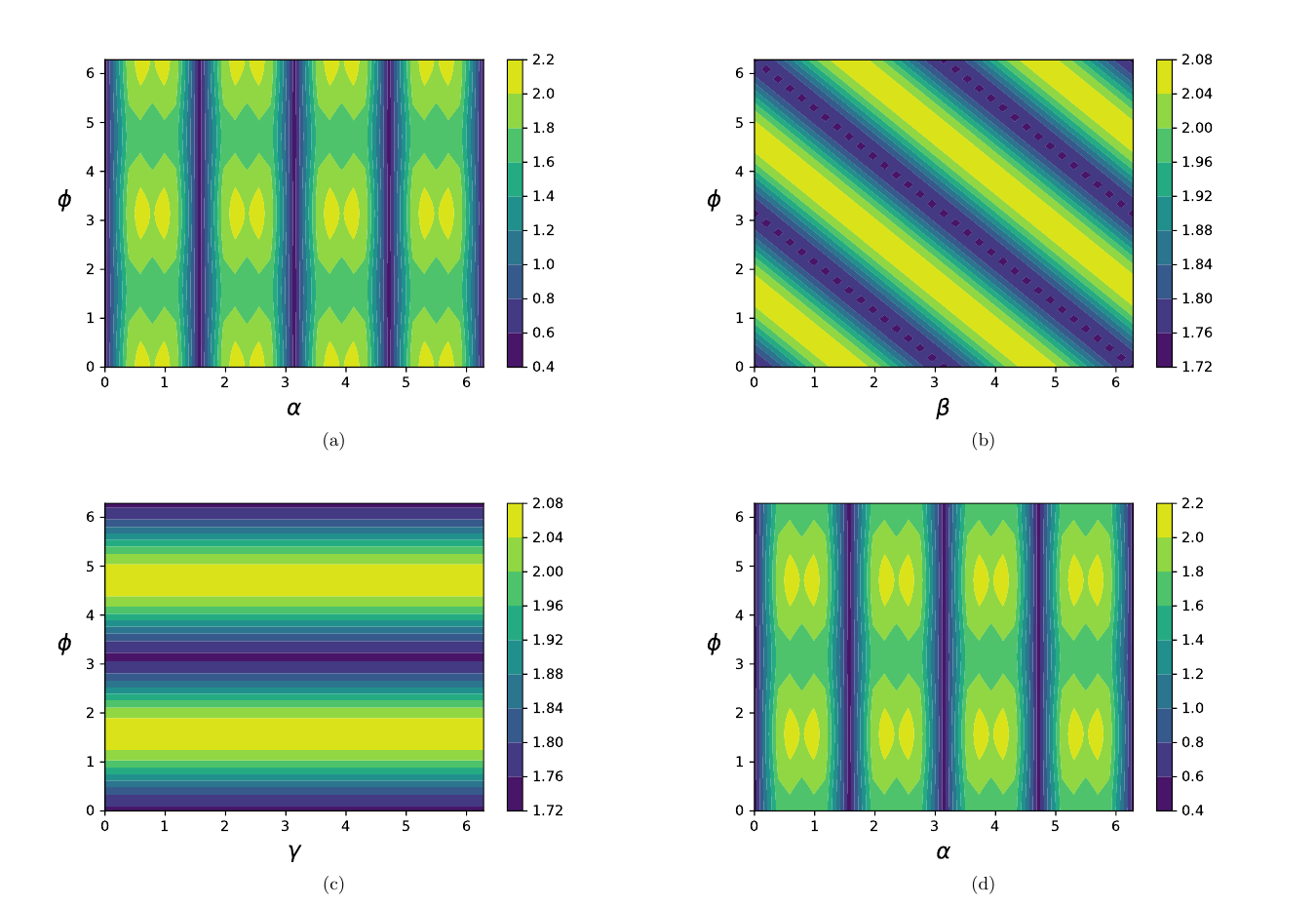}
\caption{3-way entanglement $\pi_{av}$ as a function of initial state parameter $\phi$ and coin parameters: (a) $\alpha$ with $(\beta=\frac{\pi}{2},\gamma=\frac{\pi}{2})$, (b) $\beta$ with $(\alpha=\frac{3\pi}{16},\gamma=\frac{\pi}{2})$, (c) $\gamma$ with $(\alpha=\frac{3\pi}{16},\beta=\pi)$, (d) $\alpha$ with $(\beta=\pi,\gamma=\frac{\pi}{8})$ at time step $t=2$. In all subplots, each parameter from the set $\{\phi,\alpha,\beta,\gamma\}$ is varied from 0 to $2\pi$ in steps of $\frac{2\pi}{32}$.}
\label{fs2}
\end{figure}
\section{ Simulations to achieve maximal 3-way and nonlocal 2-way entanglement via single-particle alternate 2D quantum walk}
\label{appc}
\subsection{3-way entanglement}

Herein, we discuss the numerical experiments on the average 3-way entanglement (SPE) ($\pi_{av}$) as a function of the evolution-operator and initial state parameters. The average is taken over an initial state parameter $\theta$, i.e., 
$\pi_{av}=\frac{1}{\pi}\int_{0}^{\pi}\pi_{xyc}\;d\theta$ at $t=2$. Post analysing the results shown in Fig.~\ref{fs2}, the values of the parameters at which the $\pi_{av}$ attains maximum are given in Table~\ref{t1}. One can observe that numerous optimal coin operators can be formed from Table~\ref{t1}, which yield a large $\pi$-tangle (3-way SPE). Noteworthy that increasing $t$ automatically leads to an increase in dimensions of $x$ and $y$ DoF. This causes the residual negativity values ($\pi_{x}$ and $\pi_{y}$) to increase with $t$, see Ref.~\cite{Chandra2022} and Fig.~\ref{fs1}. Investigating the QW evolution to generate 3-way SPE with a general initial state (Eq.~(\ref{eq7})) and general coin (see, Eq.~(\ref{eq8})) is computationally very costly and challenging. These two reasons lead us to optimize for the time-step $t=2$ in generating maximal 3-way SPE in the QW evolution. This is shown in Fig.~\ref{fs1}, and the results can be further utilized at large $t$ to generate maximal values of 3-way SPE ($\pi_{av}$). 

\begin{figure}[h!]
\includegraphics[width = 13cm,height=7cm]{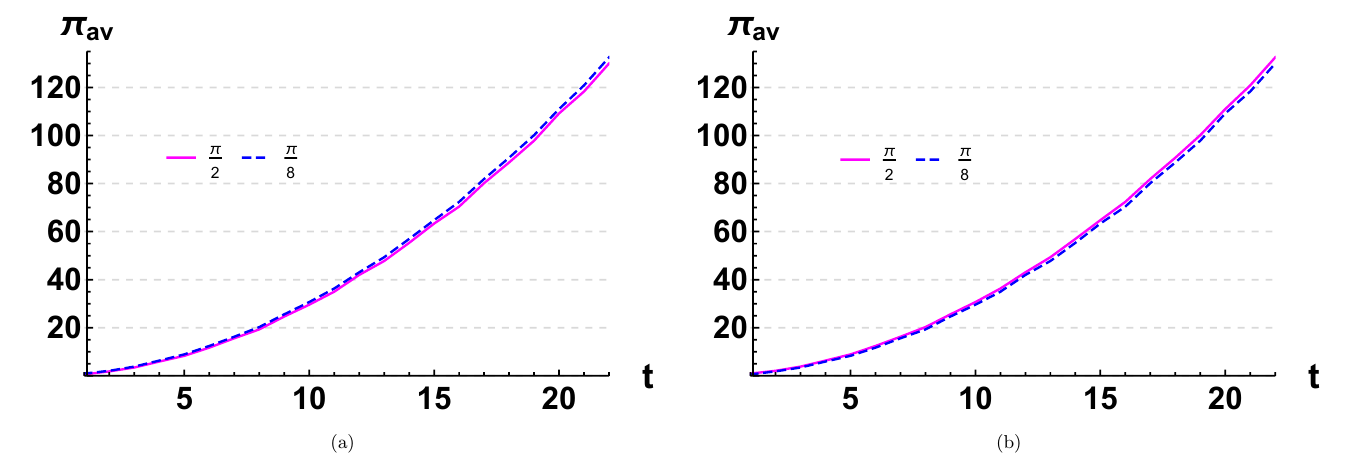}
\caption{Three-way entanglement  ($\pi_{av}$) vs time steps($t$) for spatial evolution sequences: (a) $M3_xM3_y...$ for arbitrary separable initial state Eq.~(\ref{eq7}) with $\phi=\frac{\pi}{8},\frac{\pi}{2}$; (b) $M4_xM4_y...$ for the arbitrary separable initial state with $\phi=\frac{\pi}{2},\frac{\pi}{8}$.}
\label{fs3}
\end{figure}

From the numerical results shown in Fig.~\ref{fs2}, one can form numerous optimal coins (or evolution sequences) corresponding to particular $\phi$ values, which give rise to maximum possible $\pi_{av}$ value. For example, the coins $\hat{M1}=\hat{C}(\alpha=\frac{5\pi}{16},\beta=\frac{\pi}{2},\gamma=\frac{\pi}{2})$ obtained from Fig.~\ref{fs2}(a), $\hat{M3}=\hat{C}(\alpha=\frac{5\pi}{16},\beta=\frac{6\pi}{16},\gamma=\frac{\pi}{2})$ from Fig.~\ref{fs2}(b), $\hat{M2}=\hat{C}(\alpha=\frac{5\pi}{16},\beta=\pi,\gamma=\frac{\pi}{4})$ from Fig.~\ref{fs2}(c), and $\hat{M4}=\hat{C}(\alpha=\frac{11\pi}{16},\beta=\pi,\gamma=\frac{\pi}{8})$ from Fig.~\ref{fs2}(d) yield maximum $\pi_{av}$ value, e.g., $\pi_{av}=2.0656$ at $t=2$, see Table~\ref{t1}. In the main text Fig.~2, we plotted the three-way SPE generated as a function of time steps ($t$) for the coins $\hat{M1}$ and $\hat{M2}$. In Fig.~\ref{fs3}, see 3-way SPE generated as a function of $t$ via the coins $\hat{M3}$ and $\hat{M4}$.

All sets of coin operators, along with corresponding $\phi$ values which give rise to maximum $\pi_{av}$ value, are provided in Table~\ref{t1}. 

\onecolumngrid

\begin{table}[h]
  \centering
  \begin{tabular}{|c|c|c|}
    \hline
    \textbf{Coin operator} & \textbf{Initial state phase $\phi$ and coin parameters ($\alpha,\beta,\gamma$) in units of $\frac{2\pi}{32}$} & \textbf{Maximum 3-way} \\
    \hline
    \multirow{2}{*}{$\hat{C}(\alpha,\beta=\frac{\pi}{2},\gamma=\frac{\pi}{2})$ } & $\phi=0,\alpha\in\{3,5,11,13,19,21,27,29\};\;\phi=16,\alpha\in\{3,5,11,13,19,21,27,29\}$ & \multirow{2}{*}{$\pi_{av}$=2.0656} \\
    \cline{2-2}
    & $\phi=32,\alpha\in\{3,5,11,13,19,21,27,29\}$ & \\
    \hline
    \multirow{2}{*}{$\hat{C}(\alpha=\frac{5\pi}{16},\beta,\gamma=\frac{\pi}{2})$ } & $\phi=j,\beta\in\{8-j,24-j\}|j\in[0,8]\cap\mathbb{Z};\;\phi\in\{8,24\},\beta=32$ & \multirow{2}{*}{$\pi_{av}$=2.0656} \\
    \cline{2-2}
    & $\phi=9+j,\beta\in\{31-j,15-j\}|j\in[0,15]\cap\mathbb{Z};\;\phi=25+j,\beta\in\{15-j,31-j\}|j\in[0,7]\cap\mathbb{Z}$ & \\
    \hline
    \multirow{2}{*}{$\hat{C}(\alpha=\frac{5\pi} {16},\beta=\pi,\gamma)$} & $\phi=8,\gamma\in[0,32]\cap\mathbb{Z}$  & \multirow{2}{*}{$\pi_{av}$=2.0656} \\
    \cline{2-2}
    & $\phi=24,\gamma\in[0,32]\cap\mathbb{Z}$  & \\
    \hline
   \multirow{2}{*}{$\hat{C}(\alpha,\beta=\pi,\gamma=\frac{\pi}{8})$} & $\phi=8,\alpha\in\{3,5,11,13,19,21,27,29\}$ & \multirow{2}{*}{$\pi_{av}$=2.0656} \\
    \cline{2-2}
    &$\phi=24,\alpha\in\{3,5,11,13,19,21,27,29\}$ & \\
    \hline
  \end{tabular}
  \caption{Coin operators and initial states from simulation results of Fig.~\ref{fs2}, for maximum 3-way entanglement ($\pi_{av}$).}
  \label{t1}
\end{table}

\subsection{Nonlocal 2-way entanglement}

\begin{figure*}[h]
\includegraphics[width = 18cm,height=13cm]{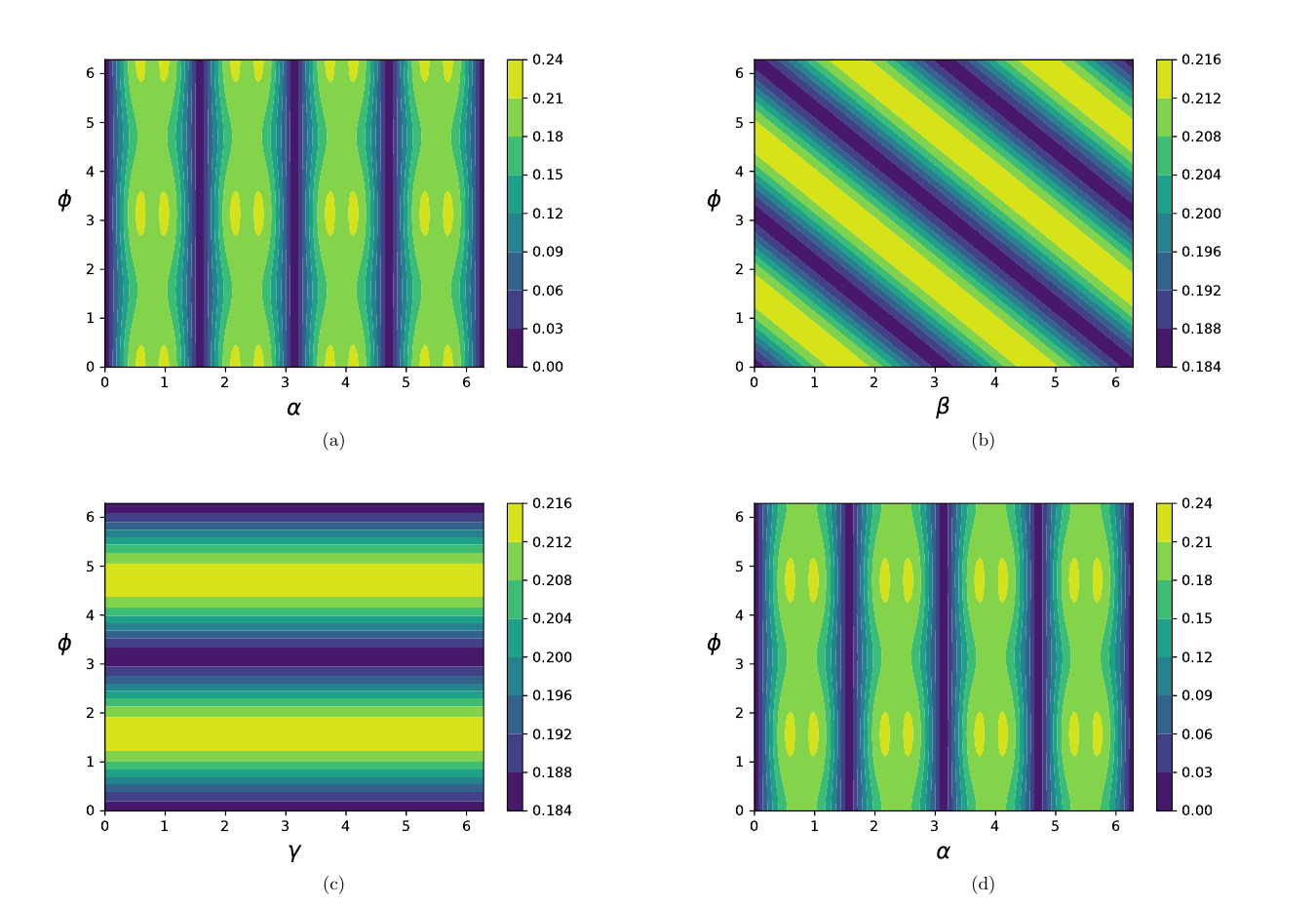}
\caption{Nonlocal entanglement ($N_{av}$) between $x,y$ DoF as a function of initial state parameter $\phi$ and coin parameters: (a) $\alpha$ with $(\beta=\frac{\pi}{2},\gamma=\frac{\pi}{2})$, (b) $\beta$ with $(\alpha=\frac{19\pi}{16},\gamma=\frac{\pi}{2})$, (c) $\gamma$ with $(\alpha=\frac{19\pi}{16},\beta=\pi)$, (d) $\alpha$ with $(\beta=\pi,\gamma=\frac{\pi}{8})$ at time step $t=2$. In all subplots, each parameter from the set $\{\phi,\alpha,\beta,\gamma\}$ is varied from 0 to $2\pi$ in steps of $\frac{2\pi}{128}$, resulting in a total 65536 combinations of separable initial states and coin operators to generate nonlocal $xy$-entanglement.}
\label{fs4}
\end{figure*}
Herein, we discuss the simulations in Fig.~\ref{fs4} to achieve a maximal value for the nonlocal 2-way single-particle entanglement between the nonlocal $x$ and $y$-DoF of the particle (e.g., the SPE between the photon’s internal OAM and its external path, i.e., the spatial trajectories). The particle evolves via the alternate 2D QW dynamics from a separable initial state given in Eq.~\ref{eq7}. The nolocal 2-way entanglement is quantified by entanglement negativity $N$, as discussed in the main (Page-3) and the average negativity, $N_{av}=\langle N \rangle = \frac{1}{\pi} \int_{0}^{\pi}\;N\;d\theta$. Pertaining to the relation given in Eqs.~(5) and (6) of main, Fig.~\ref{fs4} shows the entanglement negativity $N_{av}$ as a function of initial state parameter $\phi$ and coin parameters at time step $t=2$. Note that due to the increase of dimensions of $x$ and $y$ DoF, with an increase in QW time steps, the negativity increases with time steps ($t$), see Ref.~\cite{Chandra2022}. In addition, investigating 2D QW schemes to generate nonlocal SPE with a general initial state (Eq.~(\ref{eq7})) and general coin (see, Eq.~(\ref{eq8})) is computationally very costly and challenging. This leads us to optimize for the time-step $t=2$ in generating the nonlocal $xy$-entanglement as shown in Fig.~\ref{fs4}, and its results can be utilized for large $t$ to generate a maximal amount of entanglement as well. In all the subplots of Fig.~\ref{fs4}, each parameter from the set $\{\phi,\alpha,\beta,\gamma\}$ is varied between 0 and $2\pi$ in steps of $\frac{2\pi}{128}$, resulting in a total 65536 combinations of separable initial states and coin operators to generate nonlocal $xy$-entanglement and then determine the maximal values. In Fig.~\ref{fs4}(a), the maximization of $N_{av}$ with respect to $\phi$ and coin parameter $\alpha$ is obtained with $(\beta=\frac{\pi}{2},\gamma=\frac{\pi}{2})$. This yields a set of coin operators which with a certain initial state, generate optimum $N_{av}$ value, i.e., 0.429. Similarly, Fig.~\ref{fs4}(b) with $\beta$ $(\alpha=\frac{19\pi}{16},\gamma=\frac{\pi}{2})$, yields another set of coin operators with corresponding $\phi$ values (initial states) to generate the same optimum value $N_{av}=0.429$. Furthermore, Fig.~\ref{fs4}(c) with $(\alpha=\frac{19\pi}{16},\beta=\pi)$ give rise to another set of coin operators along with $\phi$ values to get optimum $N_{av}=0.429$. Here, it is interesting to observe that specific to any $\phi$ value, the entanglement $N_{av}$ is independent of the coin parameter $\gamma$. Fig.~\ref{fs4} (d) involves optimization with $\alpha$ at $\beta=\pi,\gamma=\frac{\pi}{8}$ and yields a large set of coin operators each of which generate maximum $N_{av}=0.429$ corresponding to certain $\phi$ values, at time step $t=2$. All these sets of coin operators along with corresponding $\phi$ values which give rise to maximum $N_{av}$ values, are provided in Table~\ref{t2}.
From Fig.~\ref{fs4} or Table~\ref{t2}, one can form numerous coin operators along with $\phi$ values (i.e., phase parameter of the initial state) to generate maximal values of entanglement between the nonlocal DoF of the particle. 

\begin{table}[h]
  \centering
  \begin{tabular}{|c|c|c|}
    \hline
    \textbf{Coin operator} & \textbf{Initial state phase $\phi$ and coin parameters ($\alpha,\beta,\gamma$) in units of $\frac{2\pi}{128}$} & \textbf{Maximum 2-way} \\
    \hline
    \multirow{2}{*}{$\hat{C}(\alpha,\beta=\frac{\pi}{2},\gamma=\frac{\pi}{2})$ } & $\phi=0,\alpha\in\{12,20,44,52,76,84,108,116\};\;\phi=64,\alpha\in\{12,20,44,52,76,84,108,116\}$ & \multirow{2}{*}{$N_{av}=0.429$} \\
    \cline{2-2}
    & $\phi=128,\alpha\in\{12,20,44,52,76,84,108,116\}$ & \\
    \hline
    \multirow{2}{*}{$\hat{C}(\alpha=\frac{19\pi}{16},\beta,\gamma=\frac{\pi}{2})$ } & $\phi=128-j,\beta\in\{96+j,32+j\}|j\in[0,32]\cap\mathbb{Z};\;\phi\in\{96,32\},\beta=0$ & \multirow{2}{*}{$N_{av}=0.429$} \\
    \cline{2-2}
    & $\phi=96-j,\beta\in\{j,64+j\}|j\in[1,64]\cap\mathbb{Z};\;\phi=32-j,\beta\in\{96+j,64+j\}|j\in[1,32]\cap\mathbb{Z}$ & \\
    \hline
    \multirow{2}{*}{$\hat{C}(\alpha=\frac{19\pi} {16},\beta=\pi,\gamma)$} & $\phi=32,\gamma\in[0,128]\cap\mathbb{Z}$ & \multirow{2}{*}{$N_{av}=0.429$} \\
    \cline{2-2}
    &$\phi=96,\gamma\in[0,128]\cap\mathbb{Z}$  & \\
    \hline
   \multirow{2}{*}{$\hat{C}(\alpha,\beta=\pi,\gamma=\frac{\pi}{8})$} & $\phi=32,\alpha\in\{12,20,44,52,76,84,108,116\}$ & \multirow{2}{*}{$N_{av}=0.429$} \\
    \cline{2-2}
    & $\phi=96,\alpha\in\{12,20,44,52,76,84,108,116\}$ & \\
    \hline
  \end{tabular}
  \caption{Coin operators and initial states from simulation results of Fig.~\ref{fs2}, for maximum 2-way nonlocal ($xy$) entanglement.}
  \label{t2}
\end{table}

For instance, we design the coin operators, 
$\hat{G1}=\hat{C}(\alpha=\frac{19\pi}{16},\beta=\frac{\pi}{2},\gamma=\frac{\pi}{2})$ with $\phi=\pi,0$ from Fig.~\ref{fs4}(a), $\hat{G3}=\hat{C}(\alpha=\frac{19\pi}{16},\beta=\frac{11\pi}{8},\gamma=\frac{\pi}{2})$ with $\phi=\frac{\pi}{8},\frac{9\pi}{8}$ from Fig.~\ref{fs4}(b), 
$\hat{G2}=\hat{C}(\alpha=\frac{19\pi}{16},\beta=\pi,\gamma=\frac{\pi}{16})$ with $\phi=\frac{\pi}{2},\frac{3\pi}{2}$ 
 from Fig.~\ref{fs4}(c), 
and 
$\hat{G4}=\hat{C}(\alpha=\frac{5\pi}{16},\beta=\pi,\gamma=\frac{\pi}{8})$ with $\phi=\frac{\pi}{2},\frac{3\pi}{2}$ from Fig.~\ref{fs4}(d) which yield maximum possible nonlocal 2-way SPE. We denote the evolution operators (Eq.~(\ref{eq8})) involving these coins to be respectively $G1_xG1_y,\;G3_xG3_y,\;G2_xG2_y$ and $G4_xG4_y$. 

\begin{figure}[h!]
\includegraphics[width = 13cm,height=7cm]{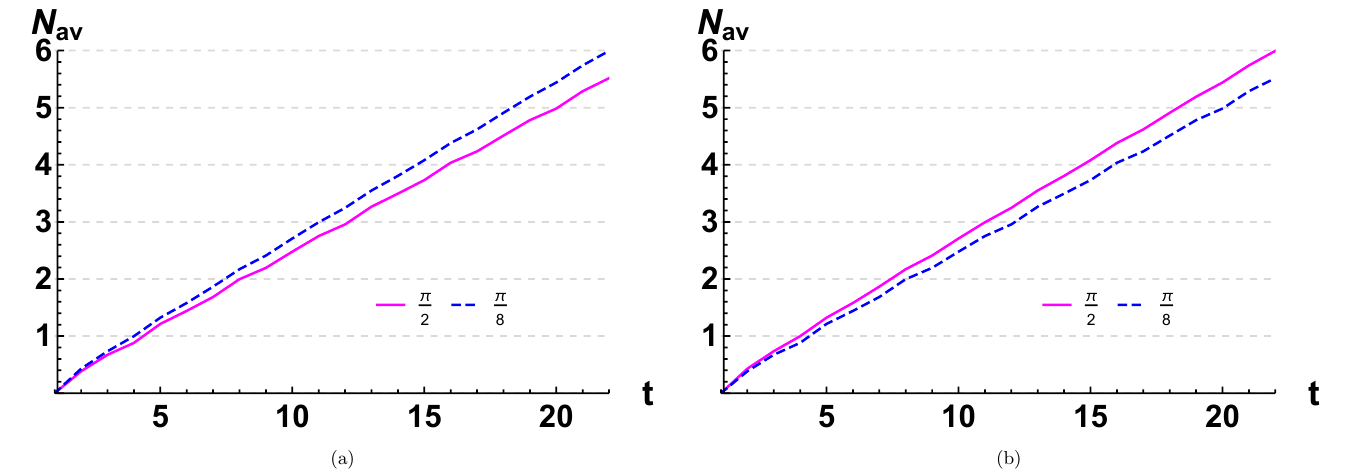}
\caption{Nonlocal entanglement ($N_{av}$) between $x,y$ DoF  vs time steps($t$) for spatial evolution sequences: $G3_xG3_y...$ for separable initial states (Eq.~(\ref{eq7})) with $\phi\in\{\frac{\pi}{2},\frac{\pi}{8}\}$ and $G4_xG4_y...$ for the separable initial states with $\phi\in\{\frac{\pi}{2},\frac{\pi}{8}\}$.}
\label{fs5}
\end{figure}

These spatial evolution sequences $G1_xG1_y,\;G3_xG3_y,\;G2_xG2_y$ and $G4_xG4_y$, yield maximum possible value of average entanglement negativity $N_{av}$ (as per Fig.~\ref{fs4}). See main Fig.~4 for 2-way SPE generation via the evolutions: $G1_xG1_y,\;G2_xG2_y$ and Fig.~\ref{fs5} below for 2-way SPE generation via the evolutions: $G3_xG3_x,\;G4_xG4_y$. The spatial evolution sequences $G3_xG3_x...,\;G4_xG4_y...$ yield maximal average entanglement negativity respectively for $\phi=\frac{\pi}{8}$ and $\phi=\frac{\pi}{2}$, as one expects from the simulations (Fig.~\ref{fs4}).

\section{Comparison between single-particle entanglement generation approaches}
\label{appd}

\begin{figure}[H]
\centering
\includegraphics[width = 13cm,height=7cm]{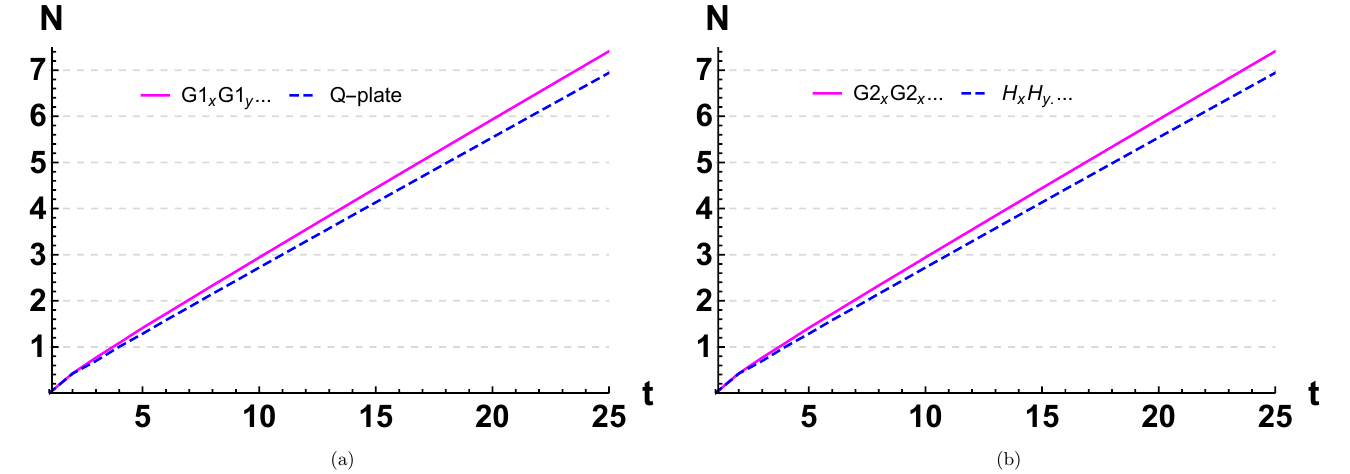}
\caption{Nonlocal entanglement ($N_{xy}$) between $x,y$ DoF, up to 25 time steps ($t$), (a) for the spatial evolution sequence $G1_xG1_y...$ vs. optimal entangler of Ref.~\cite{Chandra2022} (i.e., the q-plate coin evolution sequence) for initial state (Eq.~(\ref{eq7})) with ($\theta=\frac{\pi}{2},\phi=0$); (b) for the spatial evolution sequence $G2_xG2_y...$ vs. optimal entangler of Refs.~\cite{BuschPRA, BuschPRL} (i.e., $H_xH_y$ sequence), for initial state (Eq.~(\ref{eq7})) with ($\theta=\frac{\pi}{2},\phi=\frac{\pi}{2}$). Note that at $t=1$, $N_{av}=0$ for all four evolution sequences.}
\label{fs6}
\end{figure}

It is to be noted that there is no report of any work on generating genuine 3-way entanglement (SPE) prior to this paper. Apart from generating 3-way SPE states, this paper reports the maximum possible values for the 3-way entanglement in a single quantum particle evolving via alternate 2D QW. We also achieve maximum possible nonlocal 2-way SPE for the particle. We now compare our results on nonlocal 2-way SPE with previous relevant works~\cite{Chandra2022,BuschPRL, BuschPRA}. Ref.~\cite{BuschPRA} considers a general initial state but only with a Hadamard coin to achieve the 2-way SPE generation. On the other hand, Ref.~\cite{Chandra2022} discusses this considering a particular initial state with $\theta=\frac{\pi}{2}$ and $\phi=0$ , where each step in the QW involves the operation of a coin and its hermitian conjugate. Ref.~\cite{BuschPRL} uses Hadamard coin along with a particular initial state with $\theta=\frac{\pi}{2}$ and $\phi=\pi$ to generate SPE. Our work provides a complete investigation of 2-way (and genuine 3-way) SPE generation considering both general initial state as well as arbitrary single-qubit coin operators. Our investigation holds great importance as the QW dynamics is highly dependent on both the choices of the initial state and the coin parameters and so is the SPE generation via the QW. Fig.~\ref{fs6} shows the entanglement generated via the best entanglers from Refs.~\cite{Chandra2022,BuschPRL, BuschPRA} vs. two of our optimal entanglers ($G1_xG1_x...$, $G2_xG2_x...$). One can clearly observe that our scheme outperforms these schemes in generating nonlocal 2-way SPE at all time steps ($t$). In particular, at time-step $t=25$, our scheme exceeds both the existing schemes. Moreover, our work, apart from giving a most general investigation, provides a more resource-saving avenue to generate SPE than Ref.~\cite{Chandra2022}. The latter involves the use of a coin and its hermitian conjugate at each time step of the alternate QW, in contrary to our use of just a single coin. The similarities and comparisons between our work and the three relevant works are juxtaposed in Table~\ref{tab}. 

\begin{table*}[h!]
\centering
\caption{Comparison between nonlocal single-particle entanglement generating approaches}
\label{tab7}
\resizebox{\textwidth}{!}{%
\begin{tabular}{|c|l|l|l|l|}
\hline
\textbf{Properties$\downarrow$/Model$\rightarrow$} &
\multicolumn{1}{c|}{\textbf{\begin{tabular}[c]{@{}c@{}} Alternate QW with\\ spatial sequences\\ (This paper) \end{tabular}}} &
\multicolumn{1}{c|}{\textbf{\begin{tabular}[c]{@{}c@{}} Regular QW and\\ Hadamard-coin \\alternate QW~\cite{BuschPRL}\end{tabular}}} &
\multicolumn{1}{c|}{\textbf{\begin{tabular}[c]{@{}c@{}}Alternate QW with\\ Hadamard coin~\cite{BuschPRA}\end{tabular}}} &
\multicolumn{1}{c|}{\textbf{\begin{tabular}[c]{@{}c@{}}\\Modified Pauli \\QW Ref.~\cite{Chandra2022}\end{tabular}}} \\ \hline
\hline
\textbf{\begin{tabular}[c]{@{}c@{}}Genuine 3-way \\entanglement generation
\end{tabular}} &
\begin{tabular}[c]{@{}l@{}}Yes, and we obtain \\maximal values via\\ alternate 2D QW.
\end{tabular} &
\begin{tabular}[c]{@{}l@{}}Not attempted.\end{tabular} &
\begin{tabular}[c]{@{}l@{}}Not attempted.\end{tabular} &
\begin{tabular}[c]{@{}l@{}}Not attempted.\end{tabular} \\
\hline
\textbf{\begin{tabular}[c]{@{}c@{}}Controlled nonlocal $xy$\\entanglement Generation
\end{tabular}} &
\begin{tabular}[c]{@{}l@{}}Yes.
\end{tabular} &
\begin{tabular}[c]{@{}l@{}}Not discussed.\end{tabular} &
\begin{tabular}[c]{@{}l@{}}Not achieved.\end{tabular} &
\begin{tabular}[c]{@{}l@{}}Not achieved.\end{tabular} \\ \hline
\textbf{\begin{tabular}[c]{@{}c@{}}Maximizing nonlocal SPE ($N$) \\over initial state\\ parameters ($\theta,\phi$) \end{tabular}} &
\begin{tabular}[c]{@{}l@{}}Yes, and we average results\\ over $\theta$ values. Arbitrary coin\\ operators are considered.\end{tabular} 
&
\begin{tabular}[c]{@{}l@{}}Not discussed.\end{tabular} &
\begin{tabular}[c]{@{}l@{}}Yes, but with\\ only Hadamard coin\\(less than maximal).\end{tabular} &
Not discussed. \\ \hline
\textbf{\begin{tabular}[c]{@{}c@{}}Maximizing nonlocal SPE ($N$)\\ over coin parameters ($\alpha,\beta,\eta$)
\end{tabular} }&
\begin{tabular}[c]{@{}l@{}}Yes. \end{tabular} &
\begin{tabular}[c]{@{}l@{}} Not discussed. \end{tabular} &
\begin{tabular}[c]{@{}l@{}}Not discussed. \end{tabular} &
\begin{tabular}[c]{@{}l@{}}Partially discussed with\\ a specific initial state.\\ \end{tabular} \\ \hline
\textbf{\begin{tabular}[c]{@{}c@{}} Achieved nonlocal SPE\\ (negativity) values \end{tabular} }&
\begin{tabular}[c]{@{}l@{}}Large, e.g.,\\
$N=7.4104$ $\rightarrow$ $G1_xG1_y...$\\ at $(t=25,\phi=0,\theta=\frac{\pi}{2})$\\$N=7.4104$ $\rightarrow$ $G2_xG2_y...$\\ at $(t=25,\phi=\frac{\pi}{2},\theta=\frac{\pi}{2})$\\
(Similar results via arbitrary\\ coin operators from Table~\ref{t2}).\end{tabular} &
\begin{tabular}[c]{@{}l@{}}Best result is \\
$N=6.9429$$\rightarrow$ $H_xH_y...$ \\at $(t=25,\phi=\frac{\pi}{2},\theta=\frac{\pi}{2})$. \end{tabular}&
\begin{tabular}[c]{@{}l@{}} Best result is \\
$N=6.9429$$\rightarrow$ $H_xH_y...$ \\at $(t=25,\phi=\frac{\pi}{2},\theta=\frac{\pi}{2})$.\end{tabular} &
\begin{tabular}[c]{@{}l@{}} Best result is,\\
$N=6.9429$$\rightarrow$$C_xC_y...$ \\at $(t=25,\phi=0,\theta=\frac{\pi}{2})$.
\end{tabular}\\ \hline
\end{tabular}%
}
\label{tab}
\end{table*}

\newpage
\section{Python Code}
\label{appe}
Herein, we give a typical Python code that generates the Fig.~\ref{fs6}(a) ($G1_xG1_y$... sequence, shown in solid magenta) or any figure of the manuscript for curious researchers.

\begin{verbatim}
#Code with Python and qutip:
from numpy import *
import numpy as np
import math
import random
from scipy import integrate
from qutip import partial_transpose, Qobj
N =25 #---->
DC=np.zeros(N)

# Pre-defs :
pi=np.pi
cos=np.cos
sin=np.sin
sqrt=np.sqrt
eye=np.eye
roll=np.roll
kron=np.kron
zeros=np.zeros
exp=np.exp
empty=np.empty
outer=np.outer
trans=np.transpose
ptrace=Qobj.ptrace
phi=pi
tcoin0=np.array([[1],[0]])
tcoin1=np.array([[0],[1]])
b0=1/sqrt(2)
b1=(1j*1)/sqrt(2)

#Needed stuffs :
coin0 = np.array([1, 0]) # |0>
coin1 = np.array([0, 1]) # |1>
C00 = np.outer(coin0, coin0) # |0><0|
C01 = np.outer(coin0, coin1) # |0><1|
C10 = np.outer(coin1, coin0) # |1><0|
C11 = np.outer(coin1, coin1) # |1><1|
C_hat = (cos(pi/4)*C00 + sin(pi/4)*C01 + sin(pi/4)*C10 - cos(pi/4)*C11)
def Gencoin(th,bt,gma): 
        empe = (cos(th)*C00 + exp(1j*bt)*sin(th)*C01+exp(1j*gma)*sin(th)*C10-exp(1j*(gma+bt))*cos(th)*C11) 
        return empe
G1=Gencoin((76*2*pi)/128,pi/2,pi/2)
#----------------------------------------
def negat(l,m,phi,P): #Calc.
        posn0 = zeros(P)
        posn0[m] = 1
        psi0 = kron(posn0,kron(posn0,(cos(l/2)*coin0 + exp(1j*phi)*sin(l/2)*coin1)))
        psiN = psi0
        Plus =roll(eye(P), 1, axis=0)
        Minus = roll(eye(P), -1, axis=0)
        S_hatY = kron(kron(eye(P),Plus), C11) + kron(kron(eye(P),Minus), C00)
        S_hatX = kron(kron(Plus,eye(P)), C11) + kron(kron(Minus,eye(P)), C00)       
        for i in range(1, m + 1):
            if (i)%2==0:
              Hy = S_hatY.dot(kron(kron(eye(P),eye(P)), G1)) 
            else:
              Hy = S_hatY.dot(kron(kron(eye(P),eye(P)),G1)) 
            if i%2==0:
              Hx = S_hatX.dot(kron(kron(eye(P),eye(P)),G1))  
            else:
              Hx = S_hatX.dot(kron(kron(eye(P),eye(P)),G1))              
            psiN = np.linalg.matrix_power(Hx,1).dot(psiN) 
            psiN = np.linalg.matrix_power(Hy,1).dot(psiN)            
        DpsiN=psiN.conjugate()
        rho=outer(psiN,DpsiN)
        rho_p =(kron(kron(eye(P),eye(P)),coin0).dot(rho)).dot(kron(kron(eye(P),eye(P)),tcoin0))+
        (kron(kron(eye(P),eye(P)),coin1).dot(rho)).dot(kron(kron(eye(P),eye(P)),tcoin1))        
        PT_rho_p = partial_transpose(Qobj(rho_p,dims=[[P,P],[P,P]]), [1,0])       
        Q=Qobj(PT_rho_p,dims=[[P,P],[P,P]],isherm = True)
        eigen_values_PT=Q.eigenenergies(sparse=False,sort='low')        
        Neg=0
        for n in range(0,P**2):
          Neg=Neg+(abs(eigen_values_PT[n])-eigen_values_PT[n])/2          
        negty=Neg
        return negty
#.........................:
for m in range(1, N + 1):
    P = 2*m+1    
    l=pi/2
    ans=negat(l,m,phi,P)
    DC[m-1] = ans
    print(DC[m-1])   
print(DC)
\end{verbatim}


\twocolumngrid

\end{document}